\documentclass{aa}
\usepackage{graphicx,amssymb,psfig,epsfig}
\usepackage[authoryear]{natbib}
\begin{document}
\bibliographystyle{natbib}

\title{Evaluating GAIA performances on eclipsing binaries.}
\subtitle{III. Orbits and stellar parameters for UW~LMi, V432~Aur and CN~Lyn}

\author{P.M. Marrese\inst{1,2}
\and   U. Munari\inst{1,3}
\and   A. Siviero\inst{2}
\and   E.F. Milone\inst{4}
\and   T. Zwitter\inst{5}
\and   T. Tomov\inst{6}
\and   F. Boschi\inst{1}
\and\\ C. Boeche\inst{1}
       }
\offprints{U.Munari}

\institute {
Osservatorio Astronomico di Padova, Sede di Asiago, I-36012 Asiago (VI), Italy
\and
Dipartimento di Astronomia dell'Universit\`a di Padova, Osservatorio Astrofisico, I-36012 Asiago (VI), Italy
\and
CISAS, Centro Interdipartimentale Studi ed Attivit\`a Spaziali dell'Universit\`a di Padova, Italy
\and
Physics and Astronomy Department, University of Calgary, Calgary T2N 1N4, Canada
\and
University of Ljubljana, Department of Physics, Jadranska 19, 1000 Ljubljana, Slovenia
\and
Centre for Astronomy, Nicholaus Copernicus University, ul. Gagarina 11, 87-100 Torun, Poland
}
\date{Received date..............; accepted date................}

\abstract{
The orbits and physical parameters of three detached F and G-type eclipsing
binaries have been derived combining Hipparcos $H_{\rm P}$ photometry with
8480-8740 \AA\ ground-based spectroscopy, simulating the
photometric+spectroscopic observations that the GAIA mission will obtain. Tycho
$B_{\rm T}$ and $V_{\rm T}$ light curves are too noisy to be modeled for the three targets,
and only mean Tycho colors are retained to constrain the temperature. No previous combined
photometric+spectroscopic solution exists in literature for any of the three targets.
Quite remarkably, CN~Lyn turned
out to be an equal masses F5 triple system. Distances from the orbital solutions
agree within the astrometric error with the Hipparcos parallaxes.

\keywords{surveys:GAIA -- stars:fundamental parameters -- binaries:eclipsing --
binaries:spectroscopic}
}
\maketitle

\section{Introduction}

The Cornerstone mission GAIA, approved by ESA for launch between 2010 and
2012, will provide an astrometric, photometric and spectroscopic
all-sky survey with completeness limits for astrometry and photometry set to
$V=20$ mag (about 1 billion stars) and to $V=17.5$ mag for spectroscopy.
The limits for meaningful epoch data will be proportionally brighter.
The astrophysical and technical guidelines of the mission are described in
the ESA's {\sl Concept and Technology Study Report} (ESA SP-2000-4) and by
Gilmore et al. (1998) and Perryman et al. (2001), and in the proceedings of
recent conferences devoted to GAIA, edited by Strai\v{z}ys (1999),
Bienaym\'{e} and Turon (2002), Vansevi\'{c}ius et al. (2002) and Munari
(2003).

GAIA is expected to discover $\sim4\cdot 10^5$ eclipsing binaries,
$\sim1\cdot 10^5$ of which should be double-lined spectroscopic binaries.
This series of papers aims to ($a$) evaluate the GAIA performance on
eclipsing binaries, to the aim of providing inputs for finer tuning of
instrument focal plane assembly and data reduction pipeline, and ($b$) to
determine reasonable orbital and stellar parameters for a number of
double-lined eclipsing binaries unknown or poorly studied in literature. The
strategy we adopted to simulate GAIA observations (both photometric and
spectroscopic) is described in details in Papers~I and II (Munari et al.
2001, Zwitter et al. 2003). We briefly recall here that Hipparcos/Tycho
photometry is used as an approximation of GAIA photometry, while
ground-based spectroscopic observations (obtained with the Asiago 1.82m +
Echelle + CCD over the 8480-8740~\AA\ GAIA range) are arranged to closely
resemble GAIA spectral data.

Hipparcos scanning law and number of observations per object are pretty
close to GAIA ones, the latter however observing in more photometric
bands ($\sim$10 compared to 3). The $\sim$10 GAIA bands (the exact number
and characteristics are still subject to optimization, cf. Jordi et al.
2003) will however observe near-simultaneously during each of the $\sim$100
passages over a given star (cf. Katz 2003) during the 5~yr mission lifetime.
Therefore, they will not augment the number of points defining the
lightcurve, which will be mapped by the same $\sim$100 points as for
Hipparcos. The accuracy of the photometric solution of an eclipsing binary
rests less on the number of bands and more on the number of points mapping
the eclipses as well as the other orbital phases. Thus the Hipparcos data,
even if limited to only the $H_P$ band, still well represent the GAIA
potential to asses the stellar fundamental parameters from eclipsing
binaries.

   \begin{table*}[!t]
   \tabcolsep 0.08truecm
   \caption{Program eclipsing binaries. Data from the Hipparcos and Tycho Catalogs.
                  $H_P$ is median value from Hipparcos, $B_T$ and $V_T$ are mean values from Tycho$-$II.}
   \begin{center}
   \begin{tabular}{lccccccccccccc} \hline
   &&&&&&&&&&&&\\
   Name & & Spct. & $H_P$ & $B_T$ & $V_T$ & $\alpha_{J2000}$ & $\delta_{J2000}$ & parallax & dist & $\mu_\alpha^*$ & $\mu_\delta$ \\
        & &       &       &       &       & (h m s)          & ($^\circ$ ' ") & (mas) & (pc)& (mas yr$^{-1}$) & (mas yr$^{-1}$)\\
   &&&&&&&&&&&&\\ \hline
   &&&&&&&&&&&&\\
   UW LMi   & HIP 52465 & G0 & 8.4528 & 9.025& 8.386& 10 43 30.20 & +28 41 09.1 & 7.73$\pm$1.08 & 129$^{114}_{150}$ &$-$~~3.86$\pm$1.05 & $-$98.58$\pm$0.72 \\
   V432 Aur & HIP 26434 & G0 & 8.1377 & 8.700& 8.110& 05 37 32.51 & +37 05 12.3 & 8.43$\pm$1.58 & 119$^{100}_{146}$ & $-$39.18$\pm$1.08 & $+$34.29$\pm$0.76 \\
   CN Lyn   & HIP 39250 & F5 & 9.1026 & 9.573& 9.110& 08 01 37.20 & +38 44 58.4 & 2.76$\pm$1.53 & 362$^{233}_{813}$ &$-$~~4.43$\pm$1.91  & $+$37.80$\pm$1.03 \\
   &&&&&&&&&&&&\\
   \hline
   \end{tabular}
   \end{center}
   \end{table*}

   \begin{figure*}
   \centerline{\psfig{file=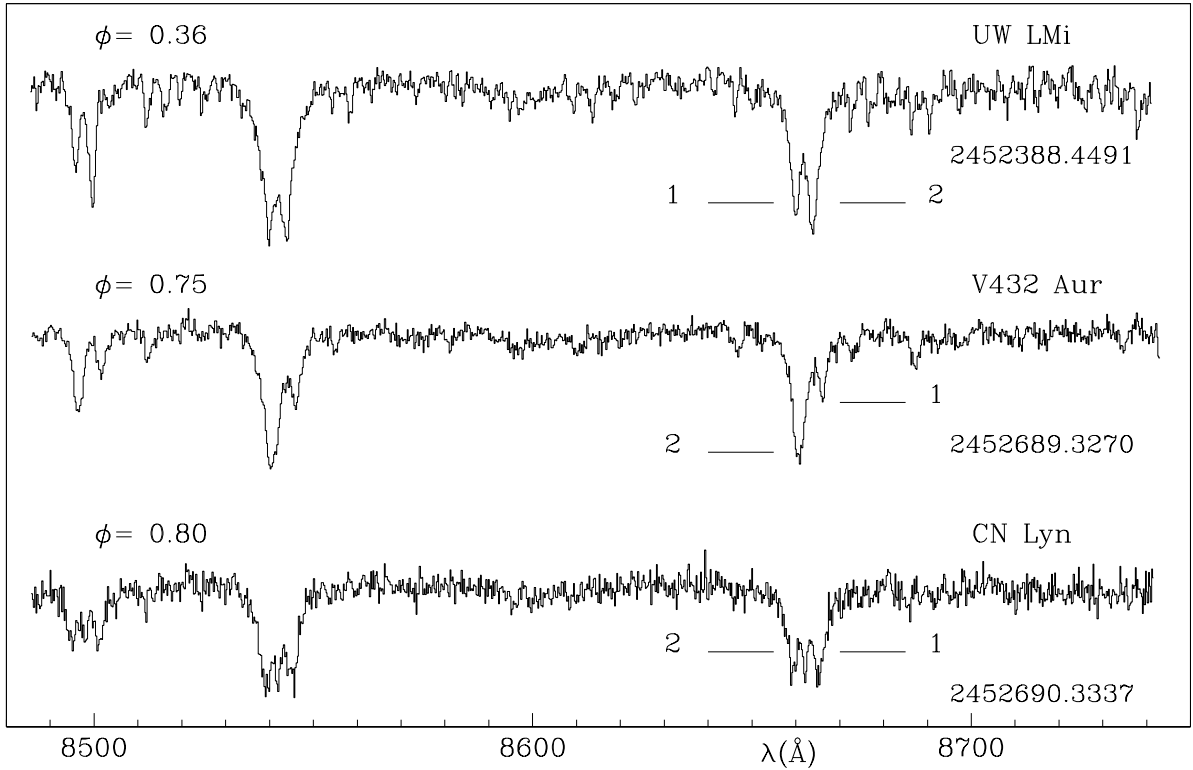,angle=270,width=16cm}}
   \caption[]{An example of normalized Asiago spectra obtained at quadratures in the GAIA
              wavelength range is shown for each target star. The numbers indicate the
              components and the HJD identify the spectra in Table~3. In UW LMi spectrum the
              intensities of the Ca~II lines are nearly equal in both components, while
              V432 Aur spectrum shows that the secondary star is more luminous. CN Lyn
              is a triple-lined system, with the central line associated to the third body.}
   \end{figure*}

   \begin{table}[!b]
   \tabcolsep 0.08truecm
   \caption{Number of Hipparcos ($H_P$) and Tycho ($B_T$, $V_T$) photometric data
   and ground based radial velocity observations, their mean S/N and standard error
   for the three program stars.}
   \begin{tabular}{lccccccccccc} \hline
   &&
   \multicolumn{2}{c}{\sl Hip}&&
   \multicolumn{3}{c}{\sl Tyc}&&
   \multicolumn{3}{c}{\sl RV}\\ \cline{3-4} \cline{6-8} \cline{10-12}
   \multicolumn{11}{c}{}\\&&
   N&$\sigma$($H_P$)&&
   N&$\sigma$($B_T$)&$\sigma$($V_T$)&&
   N&S/N&$\sigma$({\sl RV})\\
   \multicolumn{11}{c}{}\\
   UW LMi   && 110 & 0.014 && 144 & 0.13& 0.11 && 45 & 60 & 3.0 \\
   V432 Aur &&~~49 & 0.010 && ~~59& 0.10& 0.12 && 43 & 67 & 5.0 \\
   CN Lyn   &&~~69 & 0.016 && 111 & 0.18& 0.19 && 29 & 45 & 6.0 \\
   \hline
   \end{tabular}
   \end{table}

In Paper~I we selected three stars for which Tycho $B_T$ and $V_T$ light
curves provided useful constraints, while for the three stars in Paper~II
their contribution to the analysis was marginal. With present Paper~III we
push a little bit more in this direction and consider three stars for which
Tycho $B_T$ and $V_T$ light curves are useless in modeling the eclipsing
binaries, the whole analysis resting on the $H_P$ data alone. For the
program stars of this paper, only Tycho ($B_T - V_T$) mean colors are
retained and used to constrain the temperatures.

The resolving power baselined for GAIA spectrograph is $R$=11\,500 (Katz
2003), close to middle of the range 20\,000 -- 5\,000 open for evaluation by
ESA at the time this series of papers was initiated. Here we adopt a
resolving power $R$=20\,000, to balance a lower number of observations per
star (typically 30-35 vs the 100 expected from GAIA) with a better
resolution.

\section{Target selection}

For this third paper we selected three detached eclipsing binaries
poorly studied in literature, all missing a proper combined photometric +
spectroscopic solution. Their basic properties are reported in Table~1. For
the first time we were faced with the problem of determining from
spectroscopy both period and epoch of minimum, as one of the target stars
(V432 Aur) was a high amplitude variable unsolved by Hipparcos.
Table~2 summarize the number of photometric and spectroscopic data available
and their r.m.s. errors.

{\it UW~LMi}. It is a well detached eclipsing binary discovered by Hipparcos
($P$~$\sim$~$3.875$~days). It was classified as G0 V from objective prism
spectra (\citealt{upgren}). Griffin (2001) obtained a spectroscopic
solution, while Clausen et al. (2001) reported Stromgren {\em uvby} light
curves, but did not derive a photometric solution.

{\it V432~Aur}. It was recognized as a large amplitude, unsolved variable star by
Hipparcos ($H_{\rm P,min}-H_{\rm P,max}$~$\sim$~$0.38$ mag), while its
eclipsing binary nature was discovered by Dallaporta et al. (2002, hereafter D02), who
derived a period of 3.08175 days. They also discovered intrinsic variability
of the secondary star (maximum amplitude of $\Delta V$~$\sim$~$0.05$ mag),
which is not detectable in the less precise, less numerous Hipparcos data.

{\it CN~Lyn}. It is a detached eclipsing binary discovered by Hipparcos
($P$~$\sim$~$1.9554$ days). It was reported by Grenier et al. (1999) among
the stars which show a discrepancy between the luminosity class given by
visual classification and by Hipparcos parallax. Our GAIA-like spectra
reveal the system to be triple, with nearly equal luminosity components
(cf. Fig.~1).

\section{GAIA-like radial velocities and Hipparcos photometry}

The same set up as in the previous papers of this series is
maintained here. The spectra were obtained with the Echelle+CCD spectrograph
on the 1.82~m telescope operated by Osservatorio Astronomico di Padova atop
Mt. Ekar (Asiago). The spectral range covered was
$\lambda\lambda$~8480$-$8740~\AA, obtained in a single Echelle order without
gaps. As mentioned in the previous papers the actual observations cover a
wider wavelength region ($\lambda\lambda$~4550-9600~\AA), which will be
exploited elsewhere together with ground-based, multi-band, dedicated
photometry. The dispersion was 0.25~\AA/pix which, with a slit width of 2.0
arcsec, leads to a resolution of 0.42~\AA\ or equivalently to a resolving
power of $R$~=~20\,000.

The spectra were extracted and calibrated in a standard fashion using the
IRAF software package running on a PC under Linux operating system. The high
stability of the wavelength scale of the Asiago Echelle spectrograph has
been discussed in Paper I.

The radial velocity measurements along with their Heliocentric Julian Date
are given in Table~3. It is worth noticing that the smallest errors on the
radial velocities are obtained for UW~LMi which shows lines of equal
intensity for both components. Larger errors affect V432~Aur where one
component has spectral lines much weaker than the other, while the largest
errors affect CN~Lyn. The poorer S/N and the triple nature of the latter
increase the blending and decrease the line contrast against the continuum,
as the sample spectra displayed in Fig.~1 show.

Hipparcos epoch photometry was obtained from CDS. UW LMi and CN Lyn have
reliable epoch photometry (they are classified as ``class A'' in the
Hipparcos Catalogue), while V432 Aur has less reliable epoch photometry
(class C). Details on the number of observations and their accuracy are
given in Table~2.

\section{Modeling and Results}

The modeling was performed with the Wilson-Devinney code (Wilson 1998) with
modified stellar atmospheres (Milone et al. 1992) and with the newest limb
darkening coefficients (Van Hamme and Wilson 2003). Interstellar reddening
is taken to be negligible in accord with available extinction maps (Neckel
\& Klare 1980, Perry \& Johnston 1982, Burstein \& Heiles 1982).

Table~4 provides the derived system parameters along with their formal errors,
while Table~5 compares the derived distances with the astrometric
parallaxes from Hipparcos. Observational data and curves from the model
solutions are shown in Figs.~2, 3 and 4.

   \begin{table*}[!t]
   \tabcolsep 0.08truecm
   \caption{Journal of radial velocity data. The columns give the heliocentric JD
            at mid$-$exposure and the heliocentric radial velocities for both components
            (for the triple-lined system CN Lyn radial velocities refer to the
            components of the close binary).}
   \centerline{\psfig{file=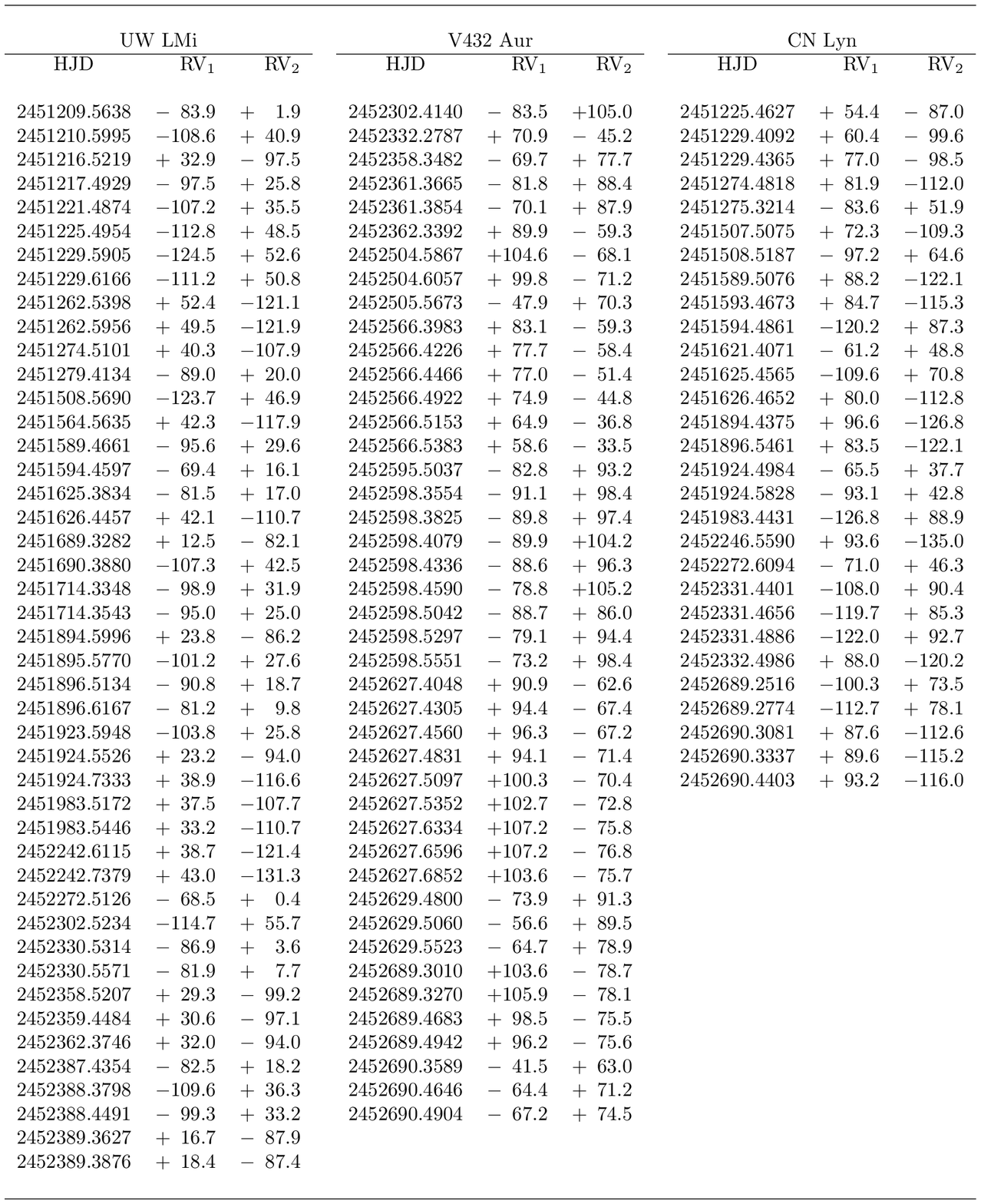,width=16.0cm}}
   \end{table*}

For all the three objects, as quoted above and as can be seen from
Figs.~2, 3 and 4, the Tycho photometric measurements are too scattered to
be used in deriving a solution. In addition, given the detached nature of
the target stars, the limited amount of points making up the $H_{\rm P}$
light curves do not properly cover both eclipses. Our solutions rely only on
$H_{\rm P}$ light curves and radial velocity curves. The usual approach to
binary star modeling is to use relative photometry obtained in each filter.
The difference of eclipse depths gives the difference of stellar
temperatures, while constraints on the absolute temperature scale can be
obtained from simultaneous solutions in different filters (a different
approach is outlined in Zwitter et al. 2003). The availability of a single
light curve does not allow us to adjust the primary temperature (i.e. obtain
the absolute temperature scale). We relied on median Tycho$-$2 colors and on
their transformation to Johnson colors to derive it. The transformation
between Tycho and Johnson systems is the same used in the previous papers,
namely $V_{\rm J}$~=~$V_{\rm T}$~$-$~$0.090$~$\times$~$(B-V)_{\rm T}$, and
$(B-V)_{\rm J}$~=~0.85~$\times$~$(B-V)_{\rm T}$ (from the Hipparcos
Catalogue). The temperatures were not adjusted during the modeling. The
temperature errors were evaluated to be $\sim$250 K. This could appear as an
overestimate, but we think it is a fair approximation as it includes
uncertainties in $B_T$ and $V_T$ measurements, their transformation to
Johnson system, color calibration of spectral types and effective
temperature calibrations of spectral types. Uncertainties in the adopted
temperatures and the poorness of phase coverage of the light curve reflect
mainly on the derived radii and luminosities and thus distances of the
target stars.

None of the target stars shows the signature of an eccentric orbit. The
circularity of the orbits was confirmed by initial modeling runs during
which eccentricity was allowed to vary and remained consistent with zero.
After a few of such trials $e$ was set to zero.

   \begin{table*}[!t]
   \caption[]{Modeling solutions. The uncertainties are formal mean
   standard errors to the solution. The last two rows give the r.m.s of the
   observed points from the derived orbital solution. $\dagger$: the
   error on the temperature is not reported because it was not adjusted in the
   modeling, it was estimated to be $\sim$250; $\ddagger$: the error on the
   temperature of the secondary component is not reported because even if it
   was adjusted in the modeling, it depends on the primary temperature, it was
   evaluated to be $\sim$260.}
   \centerline{\psfig{file=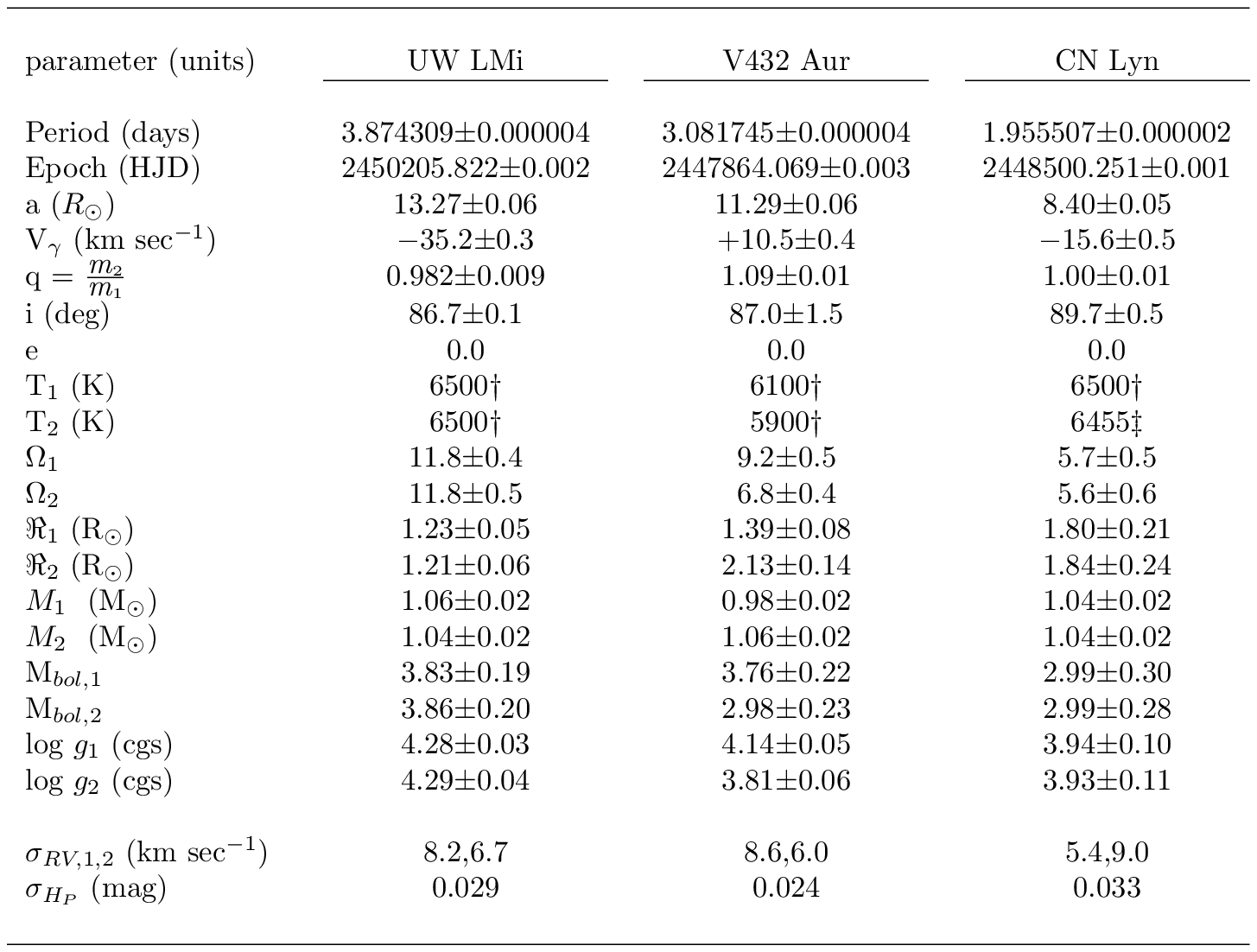,width=15cm}}
   \end{table*}
   \begin{table}[!b]
   \tabcolsep 0.08truecm
   \caption{Comparison between the Hipparcos distances and those derived
            from the parameters of the modeling solution in Table~4.}
   \begin{tabular}{lcc}
   \hline
             &Hipparcos             & this paper      \\
             & (pc)                 &  (pc)           \\
             &                      &                 \\
   UW~LMi    & 129$^{114}_{150}$    &  114$\pm$7      \\
             &                      &                 \\
   V432~Aur  & 119$^{100}_{146}$    &  124$\pm$10     \\
             &                      &                 \\
   CN~Lyn    & 362$^{233}_{813}$    &  285$\pm$32     \\
             &                      &                 \\
   \hline
   \end{tabular}
   \end{table}

\subsection{UW LMi}

For UW~LMi it has been necessary to adjust epoch and period and make the
photometric and spectroscopic ephemerids agree, because original Hipparcos
ephemeris was not working. The primary star as defined by Hipparcos is our
more massive star so we decided to retain it as primary, even if both
Griffin (2001) and Clausen et al. (2001) adopted the reverse convention. The
strategy to obtain the primary temperature was described above.
Tycho$-$2 data provide $(B-V)_{\rm T}$~=~0.625, which, transformed to the
Johnson system, leads to $(B-V)_{\rm J}$~=~0.53. An inspection of the
spectra and of the radial velocity curve tells us the two stars are not much
different. Supposing both stars have equal temperatures, the color suggests
(according to Fitzgerald 1970 and Popper 1980 conversion tables) that they
are somewhat hotter (F8) than the reported G0 spectral type. The secondary
eclipse is not well mapped in Hipparcos data and this prevented us from
adjusting even the temperature difference. Using the calibrations from
\citet{straizys1}, an F8 spectral type implies an effective temperature of
6150 K, while a G0 type gives $T_{\rm eff}$~=~5950~K. The spectrum in
Fig.~1 however shows an appreciable Paschen 14 line which intensity
relative to the CaII triplet supports a somewhat higher 6500~K temperature
(cf. synthetic spectral atlas of Munari and Castelli 2000). We thus tried
four different model solutions:

\noindent ({\sl a})~$T_{\rm eff,1}$~=~$T_{\rm eff,2}$~=~6500~K,

\noindent ({\sl b})~$T_{\rm eff,1}$~=~$T_{\rm eff,2}$~=~6150~K,

\noindent ({\sl c})~$T_{\rm eff,1}$~=~6150~K and $T_{\rm eff,2}$~=~5950~K

\noindent ({\sl d})~$T_{\rm eff,1}$~=~$T_{\rm eff,2}$~=~5950~K.

\noindent Model~{\sl a} gave a better convergence and thus $T_{\rm
eff}$~=~6500 was adopted for both stars. The relative intensity of Paschen
14 and CaII lines in the spectrum of Fig.~1 strongly support such a
temperature when compared with the synthetic spectral atlas of Munari and
Castelli (2000). The derived masses and radii ($M_{1}$~=~1.06~$M_{\odot}$,
$M_{2}$~=~1.04~$M_{\odot}$, $R_{1}$~=~1.23~$R_{\odot}$,
$R_{2}$~=~1.21~$R_{\odot}$) are consistent with those of nearly equal stars
slightly hotter and heavier than the Sun and still within the main sequence
band even if slightly away from the ZAMS locus.  Griffin (2001) suggested
that both components of the system are more luminous than MS stars. The
derived mass ratio ($q$~=~0.982$~\pm$~0.009) is completely consistent with
that obtained by Griffin (2001) whose $q$~=~1.017 transforms to $q$~=~0.983
when stars are labeled according to our scheme.

\subsection{V432 Aur}

   \begin{figure}[!t]
   \centerline{\psfig{file=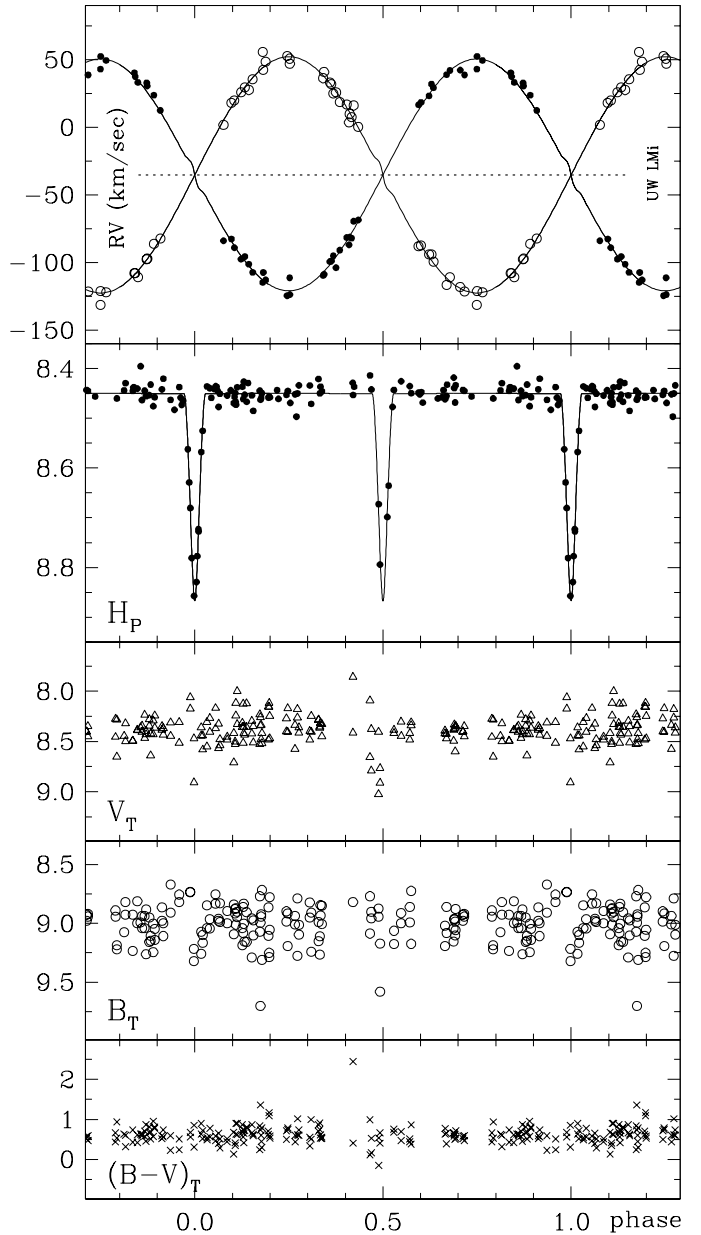,width=8.9cm}}
   \caption[]{Radial velocity curves (measurements are from Table~3, obtained in the GAIA spectral region)
              and Hipparcos $H_P$ and Tycho $V_T, B_T, (B-V)_T$ light curves
              of UW~LMi folded onto the period $P$~=~3.874309 days.
              $V_T$ and $B_T$ are shown for illustration
              purposes and were not used for modeling.
              The lines represent the solution given in Table~4, with proximity effects (appropriate if
              the co-rotation hypothesis holds) taken into account.}
   \end{figure}

   \begin{figure}[!t]
   \centerline{\psfig{file=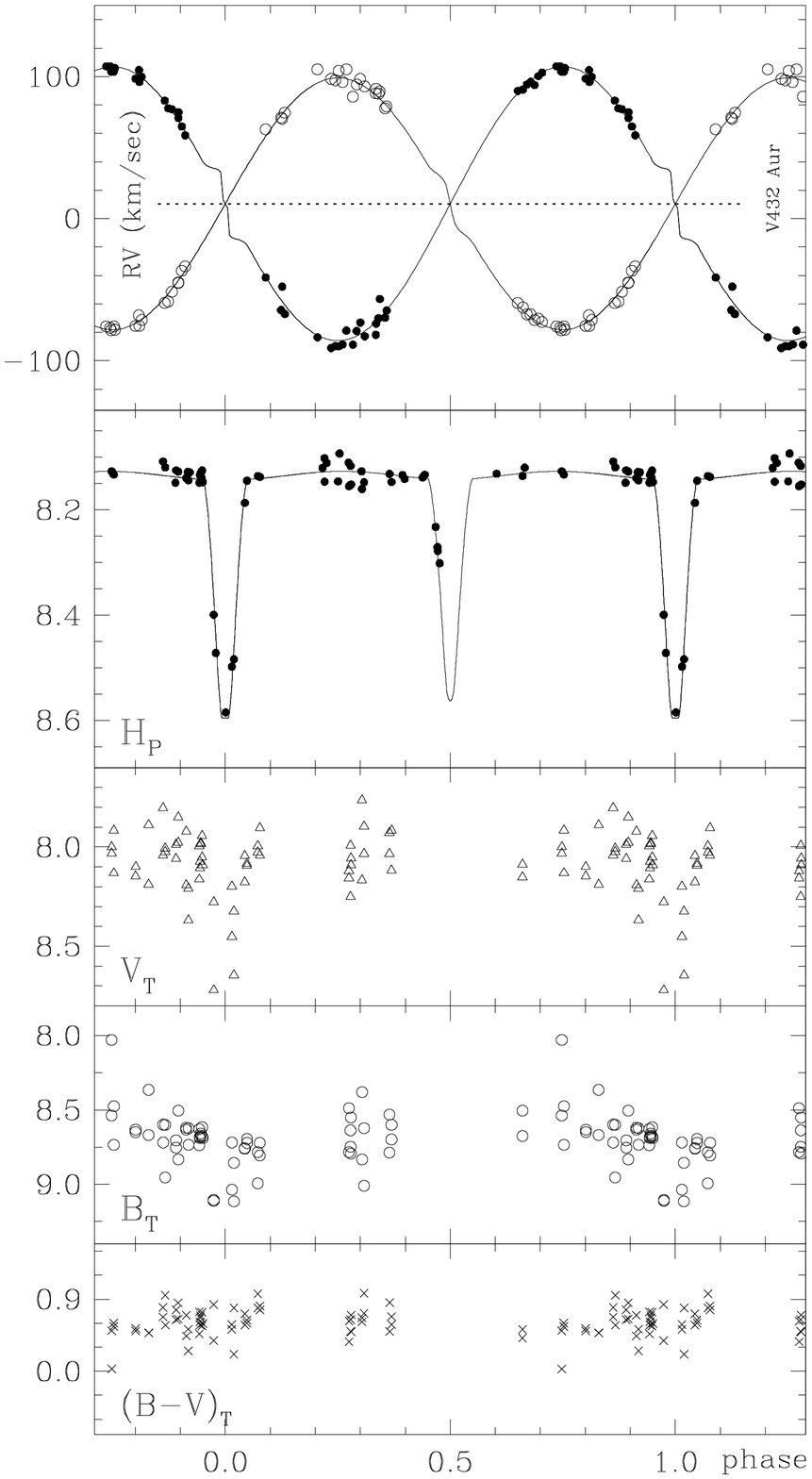,width=8.9cm}}
   \caption[]{Radial velocity curves (measurements are from Table~3, obtained in the GAIA spectral region)
              and Hipparcos $H_P$ and Tycho $V_T, B_T, (B-V)_T$ light curves
              of V432~Aur folded onto the period $P$~=~3.081745 days.
              $V_T$ and $B_T$ are shown for illustration
              purposes and were not used for modeling.
              The lines represent the solution given in Table~4, with proximity effects (appropriate if
              the co-rotation hypothesis holds) taken into account.}
   \end{figure}
   \begin{figure}[!t]
   \centerline{\psfig{file=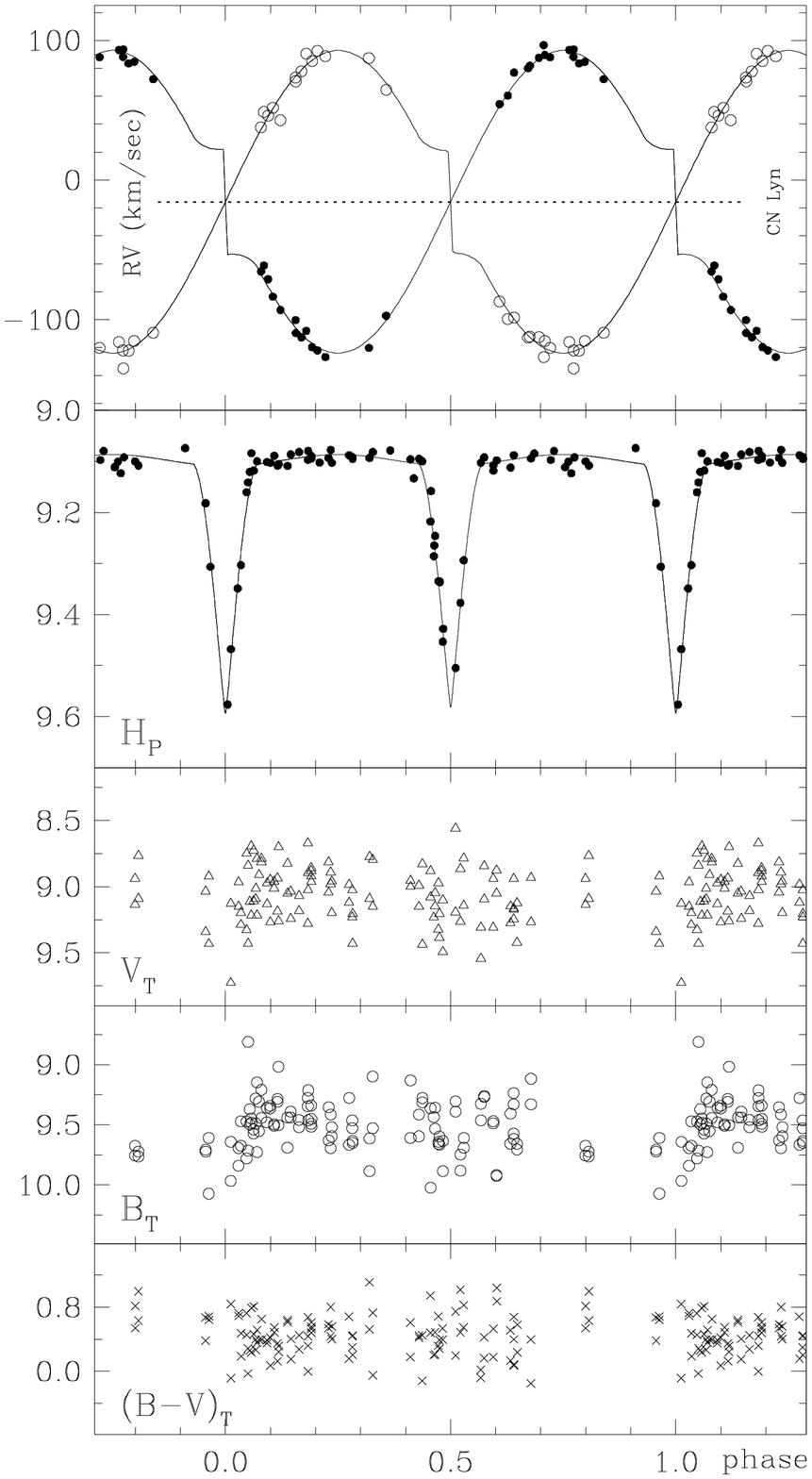,width=8.9cm}}
   \caption[]{Radial velocity curves (measurements are from Table~3, obtained in the GAIA spectral region)
              and Hipparcos $H_P$ and Tycho $V_T, B_T, (B-V)_T$ light curves
              of CN~Lyn folded onto the period $P$~=~1.955507 days.
              $V_T$ and $B_T$ are shown for illustration
              purposes and were not used for modeling.
              The lines represent the solution given in Table~4, with proximity effects (appropriate if
              the co-rotation hypothesis holds) taken into account.}
   \end{figure}

   \begin{figure}
   \centerline{\epsfig{file=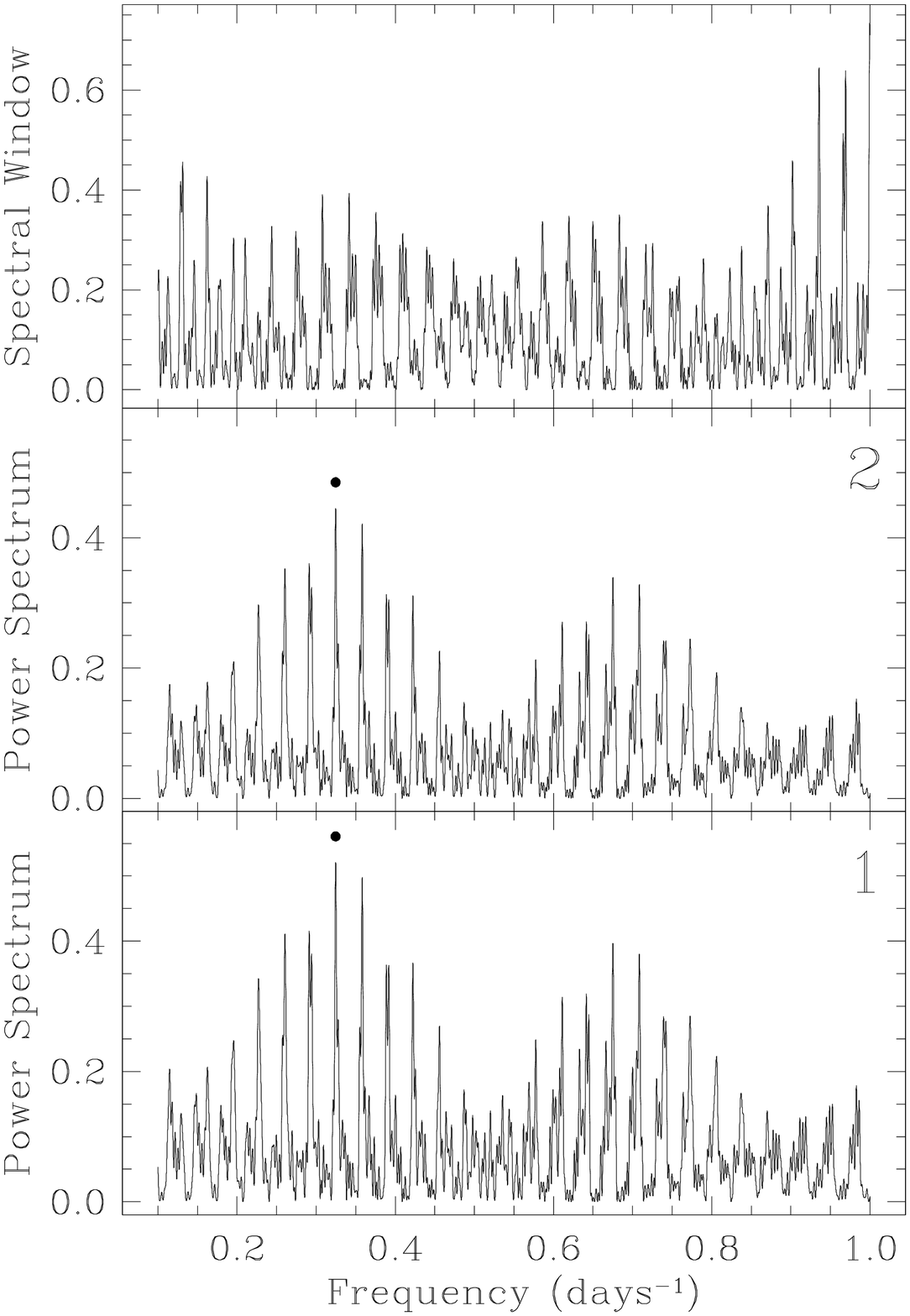,width=8.9cm}}
   \caption[]{Results of the Deeming-Fourier period search on the radial velocities
              of both V432~Aur components from Table~3. The abscissas span the 1-10 days
              interval and the dots mark the peak corresponding to the period in Table~4.
              Side peaks are the lunation aliases, while much weaker year aliases contributes
              to the background {\em noise}. The second group of peaks centered at frequency 0.65
              corresponds to half of the true period.}
   \end{figure}

V432 Aur revealed itself as the most challenging of the targets. As quoted
above, D02 derived precise a period and epoch of minimum for V432 Aur. With
the aim to rely on Hipparcos photometry and GAIA-like radial velocities
only, we neglected the D02 results and recovered the ephemeris from these data
alone. We were able to obtain initial values for period and epoch from
spectroscopy, using a Deeming-Fourier analysis that provided a clean and
fast detection of the orbital period as shown in Fig.~5. Successively, we
refined the ephemeris combining spectroscopic observations with Hipparcos
$H_{P}$ photometry.
The secondary eclipse is virtually not mapped by Hipparcos (cf. Fig.~3),
so it is highly uncertain to determine which star is the primary and to constrain
the temperature difference. As GAIA will obtain photometry in 11
photometric bands, and its spectra will be analyzed with synthetic spectral
techniques, such uncertainty will not be typical of GAIA observations of an
eclipsing system like V432~Aur, and we thus accepted to insert just
one external information, from D02, e.g. the color difference between the two eclipses.
Tycho$-$2 color $(B-V)_{\rm T}$~=~0.590 transforms to
$(B-V)_{\rm J}$~=~0.50 which, supposing both stars have equal temperatures, supports
an F7$-$F8 spectral type, somewhat hotter than the G0 type listed
in SIMBAD. The primary star is our less massive star and using $\Delta
(B-V)_{\rm J}$~=~0.05~mag at primary eclipse from D02 we obtain $\Delta T_{\rm
eff}$~$\sim$~200$-$500~K. Again we tried different models with
6000~K~$\leq$~$T_{\rm eff,1}$~$\leq$~6300~K, exploring the range 200$\leq
T_{\rm eff,1}-T_{\rm eff,2} \leq$500~K. The final adopted temperatures are
$T_{\rm eff,1}$~=~6100 and $T_{\rm eff,2}$~=~5900, corresponding to F8 and
G0 spectral types (\citealt{straizys1}), in agreement with appearance of 
the spectrum in Fig.~1. The spectra helped us to have an
initial value for the luminosity of the stars. An inspection of them
reveals that the hotter and less massive primary star is the less luminous,
because the difference in the Ca~II lines intensities cannot be ascribed to
the small temperature difference (cf. the GAIA spectral atlases by Munari \& Tomasella 1999,
and Munari \& Castelli 2000). The adopted temperatures together with the derived masses
and radii ($M_{1}$~=~0.98~$M_{\odot}$, $M_{2}$~=~1.06~$M_{\odot}$,
$R_{1}$~=~1.39~$R_{\odot}$, $R_{2}$~=~2.13~$R_{\odot}$) imply that the
secondary component is more massive, more luminous and cooler than the
primary, suggesting it has evolved farther away from the MS than the primary
component.

\subsection{CN Lyn}

This triple system is the most intriguing of the three program stars. Its
GAIA-like spectra show that the three components are quite similar in
temperature, gravity and luminosity (cf. Fig.~1). Tycho-2 photometry
$(B-V)_{\rm T}$~=~0.463 transforms to $(B-V)_{\rm J}$~=~0.39 and, supposing
all three components have equal temperatures, it suggests F3$-$F4 spectral
types, in good agreement with the F4$-$F5 type reported in SIMBAD and
the spectral appearance in Fig.~1. The adopted $T_{\rm eff,1}$ was
6500~K. Eclipses are quite well mapped by $H_{\rm P}$ photometry and the
secondary temperature was thus adjusted. The contribution of the third body
to the light curve was taken into account by adding a constant third light,
$l_{3}$~=~29($\pm$6)\% of the total flux. The derived temperatures ($T_{\rm
eff,1}$~=~6500~K, $T_{\rm eff,2}$~=~6455~K), masses
($M_{1}$~=~$M_{2}$~=~1.04~$M_{\odot}$) and radii
($R_{1}$~=~1.80~$R_{\odot}$, $R_{2}$~=~1.84~$R_{\odot}$) show that the
components of the close binary are virtually identical (within the quoted
formal errors) and marginally evolved away from the MS.

The third body mean radial velocity ($-13\pm1$~km/sec) is consistent with
the systemic velocity of the close pair
(V$_{\gamma}$~=~$-15.6\pm0.5$~km/sec). A period search has failed to reveal
a clear periodicity in the radial velocities of the third body, which
limited range of radial velocities displayed in our spectra (21~km/sec) can
be caused by an orbital period much longer than the time spanned by our
observations (3.2 yr) and/or a low orbital inclination.

\section{Conclusions}

The three targets were chosen without restrictions on peculiarities or
variability and faint enough so that Tycho~2 photometry is useful only in
providing mean colors and not epoch data. Only $H_P$ data were therefore
available to map the lightcurve, which compromised the accuracy of derived
effective temperatures of the components (both the absolute scale and the
difference).

Even under such limitations (aiming to simulate GAIA observations of targets
more difficult that those explored in Paper I and II of this series)
reasonable solutions have been achieved for all the three targets, with the
distance implied by the orbital modeling within the uncertainty bar of the
Hipparcos parallax for all the three stars, including the equal masses
triple system. This reassuring external and independent check reinforces the
expectation of a high impact of GAIA observations for the derivation of
fundamental stellar properties from observations of eclipsing binaries, and
the direct use of GAIA observations of eclipsing binaries as a valuable
measure of distances in itself.

\begin{acknowledgements}
Generous allocation of observing time with the Asiago telescopes has been
vital to this project. This research has made use of the SIMBAD database of
the Centre de Donn\'{e}es de Strasbourg. The financial support from the
Slovenian Ministry for Education, Science and Sports (to TZ), from the
Polish KBN Grant (No. 5 P03D 003 20, to TT), from the Canadian NSERC (to
EFM) and from the Italian Space Agency (contract ASI-I-R-117/01 to UM) are
kindly acknowledged.
\end{acknowledgements}


\begin{thebibliography}{}
 \bibitem[Andersen et al.(1991)]        {andersen}     Andersen, J.
                                                       1991, A\&A Review 3, 91
 \bibitem[Bienaym\`{e} \& Turon(2002)]  {bien} Bienaym\`{e} O., Turon, C. 2002, ed.s {\it GAIA: a European space project},
                                                       EAS Pub. Ser. vol. 2, EDP Sciences
 \bibitem[Burstein \& Heiles (1982)]    {bur}           Burstein, D., Heiles, C.
                                                       1982, AJ 87, 1165
 \bibitem[Clausen et al.(2001)]          {clausen}      Clausen, B. E., Helt, B. E., Olsen, E. H.
                                                       2001, A\&A 374, 980
 \bibitem[Dalla Porta et al.(2002)]      {dallap}       Dallaporta, S., Tomov, T., Zwitter, T., Munari, U.
                                                       2002, IBVS 5319 (D02)
 \bibitem[Fitzgerald(1970)]             {fitzgerald}   Fitzgerald, M. P.
                                                       1970, A\&A 4, 234
 \bibitem[Gilmore et al.(1998)]         {gilmore}      Gilmore, G., Perryman, M. A. C., Lindegren, L.
                                                       Favata, F., Hoeg, E., Lattanzi, M., Luri, X., Mignard, F., Roeser, S.,
                                                       de Zeeuw, P. T.
                                                       1998, Proc SPIE Conference 3350, p. 541
 \bibitem[Grenier et al. (1999)]         {grenier}     Grenier, S., Baylac, M.O., Rolland, L., Burnage, R., Arenou, F.,
                                                       Briot, D., Delmas, F., Duflot, M., Genty, V., G\`{o}mez, A.E., Halbwachs, J.-L.,
                                                       Marouard, M., Oblak, E., Sellier, A. 1999, A\&AS 137, 451
 \bibitem[Griffin (2001)]                {griffin}      Griffin, R. F.
                                                       2001, Obs. 121, 315
 \bibitem[Jordi et al. (2003)]          {jordi}        Jordi, C., Carrasco, J.M., Figueras, F., Torra, J. 2003, in
                                                       {\it GAIA Spectroscopy, Science and Technology}, U.Munari ed., ASP Conf. Ser. vol 298, pag. 209
 \bibitem[Katz (2003)]                  {katz}         Katz, D. 2003, in {\it GAIA Spectroscopy, Science and Technology},
                                                       U.Munari ed., ASP Conf. Ser. vol 298, pag. 65
 \bibitem[Milone et al.(1992)]          {milone}       Milone, E. F., Stagg, C.R., Kurucz, R.L.
                                                       1992, ApJS 79, 123
 \bibitem[Munari(2003)]                 {munari2}      Munari U. 2003, ed. {\it GAIA Spectroscopy, Science and Technology},
                                                       ASP Conf. Ser. vol 298, San Francisco
 \bibitem[Munari and Tomasella (1999)]  {munari3}      Munari, U., Tomasella, L. 1999, A\&A 137, 521
 \bibitem[Munari and Castelli (2000)]   {munari4}      Munari U., Castelli F. 2000, A\&AS 141, 141
 \bibitem[Munari et al.(2001)]          {munari1}      Munari, U., Tomov, T., Zwitter, T.,Milone, E. F., Kallrath, J.,
                                                       Marrese, P. M., Boschi, F., Pr\v sa, A., Tomasella, L., Moro, D.
                                                       2001, A\&A 378, 477 (Paper I)
 \bibitem[Neckel \& Klare (1980)]       {neckel}       Neckel, T., Klare, G.
                                                       1980, A\&AS 42, 251
 \bibitem[Popper(1900)]                 {popper}       Popper, D.M.
                                                       1980, ARA\&A 18, 115
 \bibitem[Perry \& Johnston (1982)]     {perry}        Perry, C.L., Johnston, L.
                                                       1982, ApJS 50, 451
 \bibitem[Perryman et al.(2001)]        {perryman}     Perryman, M.A.C., de Boer, K.S., Gilmore, G., Hoeg, E., Lattanzi, M.G.,
                                                       Lindegren, L., Luri, X., Mignard, F., Pace, O., de Zeeuw, P.T.
                                                       2001, A\&A 369, 339
 \bibitem[Stellingwerf(1978)]            {stelli}      Stellingwerf, R. F.
                                                       1978, ApJ 224, 953
 \bibitem[Strai\v{z}ys \& Kuriliene(1981)] {straizys1} Strai\v{z}ys, V., Kuriliene, G.
                                                       1981, Ap\&SS 80, 353
 \bibitem[Strai\v{z}ys(1999)]            {stra}        Strai\v{z}ys V. 1999, ed. {\it GAIA Leiden Workshop}, Baltic Astron. (special edition), vol. 8, N. 1-2
 \bibitem[Upgren \& Staron(1970)]         {upgren}     Upgren, A. R., Staron, R. T.
                                                       1970, ApJS 19, 367
 \bibitem[Van Hamme(2003)]                {vanh}       Van Hamme, W., Wilson, R.E. 2003, in GAIA Spectroscopy, Science and Technology,
                                                       U. Munari ed., ASP Conf. Series 298, p.323
 \bibitem[Vanseci\v{c}ius(2002)]          {vans}       Vanseci\v{c}ius V., Ku\v{c}inskas A., Sud\v{z}ius J. 2002, ed.s {\it Census of the
                                                       Galaxy: challeges for photometry and spectrometry with GAIA}, Kluwer
 \bibitem[Wilson(1998)]                   {wilson}     Wilson, R. E. 1998, {\it Computing Binary Star Observables},
                                                       Univ. of Florida Astronomy Dept.
 \bibitem[Zwitter (2003)]                {zwitter1}    Zwitter, T. 2003, in {\it GAIA Spectroscopy, Science and Technology},
                                                       U. Munari ed., ASP Conf. Series 298, p.329
 \bibitem[Zwitter et al.(2003)]         {zwitter2}     Zwitter, T., Munari U., Marrese, P. M., Pr\v sa, A., Milone, E. F.,
                                                       Boschi, F., Tomov, T.
                                                       2003, A\&A 404, 333 (Paper II)
\end{thebibliography}
\end{document}